\documentclass[conference]{IEEEtran}
\IEEEoverridecommandlockouts
\usepackage[top=0.75in, bottom=1.0in, left=0.625in, right=0.625in]{geometry} % ICC submission edas
\usepackage{cite}
\usepackage{amsmath,amssymb,amsfonts}
\usepackage{graphicx}
\usepackage{textcomp}
\usepackage{xcolor}
\usepackage{mathtools}
\usepackage{makecell}
\usepackage{caption}
\usepackage{subcaption} 
\usepackage{array}
\usepackage{siunitx}
\usepackage{url}

\usepackage{fancyhdr}
\fancypagestyle{firststyle}{\fancyhf{}
		\fancyhead[L]{S. Jia, M. Ying, M. Mezzavilla, T. S. Rappaport, and S. Rangan, ``Distributed Uplink Anti-Jamming in LEO Mega-Constellations via Game-Theoretic Beamforming", to appear in \textit{IEEE International Conference on Communications (ICC), }Glasgow, UK, Jun. 2026, pp. 1--6.}
	}
\usepackage{times}
\def\BibTeX{{\rm B\kern-.05em{\sc i\kern-.025em b}\kern-.08em
    T\kern-.1667em\lower.7ex\hbox{E}\kern-.125emX}}

\def\Exp{\mathbb{E}\,}

\newcommand{\bs}{\boldsymbol}
\newcommand{\mc}{\mathcal}
\newcommand{\wh}{\widehat}

\newcommand{\herm}{^{\text{\sf H}}}

\newcommand{\C}{\mathbb{C}}

\newcommand{\detm}{\operatorname{det}}

\usepackage[ruled,vlined]{algorithm2e}  % 保留折线竖线

\SetNlSty{textbf}{(}{)}           % 行号样式：(1),(2)...
\SetNlSkip{-0.45em}               % 行号再向左挪一点
\SetAlgoInsideSkip{smallskip}     % 算法上下留白稍紧凑

\SetKwInput{KwIn}{Input}
\SetKwInput{KwOut}{Output}
\SetKwInput{KwInit}{Initialize}
\SetAlgoNlRelativeSize{-1}     % 行号更细一点
\SetAlFnt{\small}              % 正文小一号，更紧凑
\SetKwComment{tcp}{// }{}      % 注释风格
\usepackage{multirow}
\SetAlFnt{\footnotesize}           % 双栏里更紧凑
\DontPrintSemicolon                % 去行尾分号（IEEE风格更干净）
\SetKwInput{KwIn}{Input}
\SetKwInput{KwOut}{Output}
\SetKwInput{KwInit}{Initialize}
\SetKwComment{tcp}{// }{}          % 注释样式

\DeclareSIUnit{\dBm}{dBm} % 兼容旧版本，若已存在也不报错

\begin{document}

\title{Distributed Uplink Anti-Jamming in LEO Mega-Constellations via Game-Theoretic Beamforming}

\IEEEoverridecommandlockouts % allow \thanks in conference mode

\author{
    \IEEEauthorblockN{
        Shizhen Jia\IEEEauthorrefmark{1},
        Mingjun Ying\IEEEauthorrefmark{1},
        Marco Mezzavilla\IEEEauthorrefmark{2},
        Theodore S. Rappaport\IEEEauthorrefmark{1},
        and Sundeep Rangan\IEEEauthorrefmark{1}
    }
    \IEEEauthorblockA{
        \IEEEauthorrefmark{1}NYU WIRELESS, New York University, Brooklyn, NY, USA\\
        \IEEEauthorrefmark{2}Dipartimento di Elettronica, Informazione e Bioingegneria (DEIB), Politecnico di Milano, Milan, Italy\\
        {\{s.jia, srangan\}@nyu.edu}
    }
    \thanks{This work was supported, in part, by NSF grants 2345139, 2148293, 2133662, 1952180, and 1904648;
    the NTIA Public Wireless Innovation Fund, and the industrial affiliates of NYU WIRELESS.
    Code is available at https://github.com/SJ00425/NTN-Anti-Jamming.}
    \vspace{-20pt}
}
\maketitle
\bstctlcite{BSTcontrol}

\thispagestyle{firststyle}

\begin{abstract}
Low-Earth-Orbit (LEO) satellite constellations have become vital in emerging commercial and defense Non-Terrestrial Networks (NTNs). However, their predictable orbital dynamics and exposed geometries make them highly susceptible to ground-based jamming. Traditional single-satellite interference mitigation techniques struggle to spatially separate desired uplink signals from nearby jammers, even with large antenna arrays. This paper explores a distributed multi-satellite anti-jamming strategy leveraging the dense connectivity and high-speed inter-satellite links of modern LEO mega-constellations. We model the uplink interference scenario as a convex-concave game between a desired terrestrial transmitter and a jammer, each optimizing their spatial covariance matrices to maximize or minimize achievable rate. We propose an efficient min–max solver combining alternating best-response updates with projected gradient descent, achieving fast convergence of beamforming strategy to the Nash equilibrium. Using realistic Starlink orbital geometries and Sionna ray-tracing simulations, we demonstrate that while close-proximity jammers can cripple single-satellite links, distributed satellite cooperation significantly enhances resilience, shifting the capacity distribution upward under strong interference.

\end{abstract}

\begin{IEEEkeywords}
LEO satellites, anti-jamming, non-terrestrial networks, game theory
\end{IEEEkeywords}

\section{Introduction}

\begin{figure}[!t]
    \centering
    \includegraphics[width=0.35\textwidth]{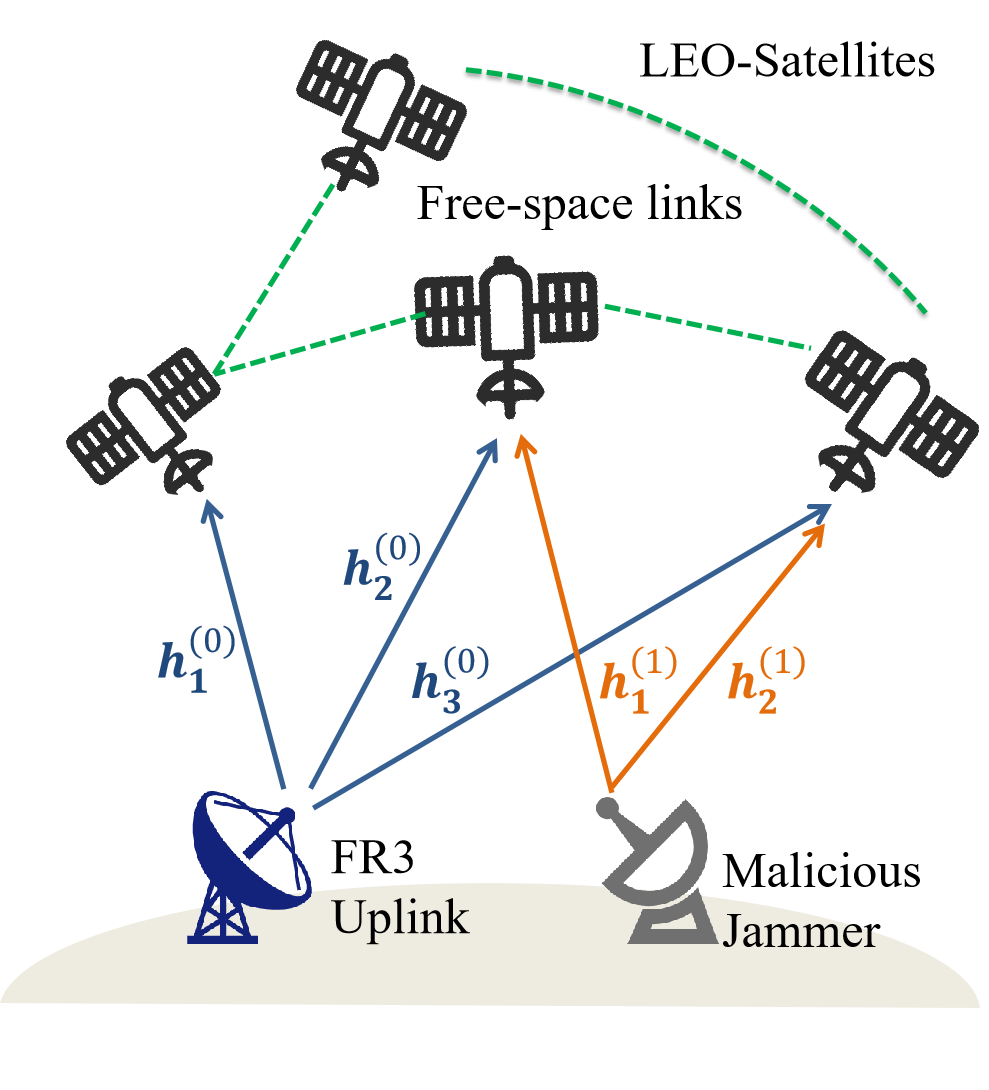}
    \vspace{-15pt}
    \caption{LEO satellite network topology with legitimate uplink and jamming channels from a malicious ground station. Inter-satellite free-space links enable cooperation among satellites.}
    \vspace{-15pt}
    \label{anti}
\end{figure}

Low-Earth-orbit (LEO) satellite systems promise truly global broadband coverage and are central to emerging Non-Terrestrial Network (NTN) architectures~\cite{rappaport2025spectrum, Xing2021CL}. However, LEO satellites' large link budgets, exposed geometries, and reliance on predictable orbital trajectories make LEO constellations acutely vulnerable to deliberate jamming~\cite{laursen2023satellite,hossein2022jamming}. 
Traditional array-based interference mitigation are typically ineffective on individual satellites:  
at large distances, ground jammers and desired transmitters are difficult to spatially separate,
even when the satellites are equipped with very large arrays.

The development of LEO mega-constellations motivates considering distributed 
multi-satellite anti-jamming.
At current densities of commercial networks such as Starlink, ground-based transmitters
can routinely see tens of satellites at reasonable path loss -- see, for example, \cite{kang2024terrestrial} and simulations herein.  Moreover, 
LEO satellite constellations are already considering high-speed laser inter-satellite links
that support data rates of  tens of Gbps, with future systems potentially 
achieving as much as a Tbps \cite{chaudhry2022temporary,wang2024free}.  These connected,
multiple satellite constellations can thus form a virtual wide-area distributed array.  
While a jammer close to a desired transmitter on the ground are too difficult to separate at individual satellite, a distributed satellite constellation can potentially null interferers
more easily.

To explore the possibility of using distributed satellites, this paper considers a simple interference scenario shown in 
Fig.~\ref{anti}.  A ground-based desired transmitter (TX$_0$) sends signals to $M$ satellites
in the presence of a ground-based jammer (TX$_1$).  The satellites are assumed to be inter-connected
so they form a space-based virtual MIMO array.  Both the desired and/or malicious jammer may have steerable arrays
so that they can modify the energy distribution to the different satellites.
The problem is how the desired transmitter should adapts its beamforming in the presence of a jammer. 

\noindent \textbf{Our Contributions:}
\begin{itemize}
\item \textbf{Game-Theoretic Formulation:} 
We formalize the uplink anti-jamming problem for distributed LEO satellites as a convex-concave game between the desired ground transmitter and a ground jammer. Specifically, the desired ground transmitter and ground jammer must select spatial covariance matrices $\bs{Q}_0$ and $\bs{Q}_1$ where they seek to maximize and minimize the uplink rate achievable by the desired ground transmitter.  It is shown that this game admits a Nash equilibrium.
\item \textbf{Efficient Min–Max Solver:} 
To find the saddle point to the min-max game, we propose a 
a lightweight alternating Best-Response + projected gradient algorithm, combining water-filling for the projected transmitter and projected gradient descent for the jammer, ensuring fast convergence to the equilibrium.
\item \textbf{LEO satellite case study and resilience analysis:} 
Using realistic Starlink orbital geometry, we evaluate both dish-type  and intelligent jammers and show that while close jammer proximity can severely degrade single-satellite links, leveraging multiple satellites significantly improves robustness and shifts the capacity distribution upward under strong interference.
\end{itemize}

% \begin{figure}[!t]
%     \centering
%     \includegraphics[width=0.38\textwidth]{plot/antijamming.png}
%     % \vspace{-15pt}
%     \caption{LEO satellite network topology with legitimate uplink and jamming channels from a malicious ground station. Inter-satellite free-space links enable cooperation among satellites.}
%     % \vspace{-10pt}
%     \label{anti}
% \end{figure}

\noindent
\textbf{Prior Work}:  Several recent works have considered distributed reception from multiple satellites to enhance coverage and capacity of the NTN network~\cite{zhang2025enabling,xu2024enhancement,zheng2022intelligent, olson2024tractable}. 
More recently, multi-satellite solutions have also been considered from improved 
spectrum usage in multi-operator scenarios~\cite{kim2025feasibility,kim2024specshar}.
Specific to jamming, 
the work \cite{kantheti2023anti} considers single source jammer (equivalent in our model to the dish-type jammer in the simulation) and provides a USRP implementation.

In MIMO processing more generally, there are a large number of works using game-theoretic interference management approaches~\cite{scutari2009mimo,scutari2008asynchronous,scutari2008competitive,zhou2014network}.
While these works treat interference as exogenous noise, we explicitly model the adversarial interaction where both players optimally adapt their strategies. We develop an alternating water-filling gradient-descent algorithm that operates directly on transmit covariance matrices and converges to an equilibrium saddle point.

A critical consideration for inter-satellite interference mitigation is that, when satellites are separated by hundreds of kilometers, per-link propagation delays vary significantly, pulse-shaping filters introduce inter-symbol interference across receivers, and independent local oscillators induce phase drift~\cite{chen2024asynchronous}. Our current analysis assumes that the links can be equalized on a per-satellite basis to avoid the propagation and frequency offsets. Addressing these effects in a manner similar to \cite{chen2024asynchronous} can be an avenue of future work.

\section{Problem Formulation}

We consider the jamming scenario in Fig.~\ref{anti}.  There are 
two ground-based transmitters:  A desired transmitter, TX$_0$, and
jamming transmitter, TX$_1$.  The desired and jamming transmitters send uplink signals to $M$
satellites.  The transmitters TX$_0$ and TX$_1$ have $N_0$ and $N_1$ elements each,
and each satellite $m$ has $K_m$ elements.
We assume that the received signal each satellite $m$ can be modeled as a 
memoryless channel:
\begin{equation} \label{eq:rm}
    \bs{r}_m[k] = \bs{H}_{0,m}\bs{x}_0[k] + \bs{H}_{1,m}\bs{x}_1[k] + \bs{d}_m[k]
    \in \C^{K_m},
\end{equation}
where $\bs{x}_0[k]$ and $\bs{x}_1[k]$ are the TX symbol vectors from the desired and interfering sources and $\bs{d}_m[k]$ and $\bs{H}_{i,m}$ are channel matrices.
Importantly, the channel matrices include the antenna gains at both the satellite and transmitters -- so we are implicitly assuming that the mechanical orientation of the transmitters are fixed.
We model the noise as AWGN:
\begin{equation}
    \bs{d}_m[k] \sim \mc{CN}(\bs{0}, \kappa \bs{I}),
\end{equation}
and $\kappa$ is the noise energy per symbol.  We use the notation $\kappa$ to not confuse
the value with $N_0$ -- the number of TX elements for TX$_0$.

We let $\bs{Q}_0$ and $\bs{Q}_1$ denote the TX covariance matrices of TX$_0$ and TX$_1$, respectively:
\begin{equation} \label{eq:Q01}
    E[\bs{x}_0[k] \bs{x}_0\herm[k]]= \bs{Q}_0, 
    \quad 
    E[\bs{x}_1[k] \bs{x}_1\herm[k]]= \bs{Q}_1.
\end{equation}
We assume that they are bounded as:
\begin{equation}
    \mathrm{Tr}(\bs{Q}_0) \leq \mc{E}_0,
    \quad
    \mathrm{Tr}(\bs{Q}_1) \leq \mc{E}_1,
\end{equation}
where $\mc{E}_0$ and $\mc{E}_1$ represent the energies per symbol.

The memoryless model \eqref{eq:rm} requires some justification.
First, consider the signal, $\bs{x}_0[k]$, from the desired transmitter, TX$_0$.
In general, due to the large distances between satellites, 
the delays from TX$_0$ to the $M$ satellites can be significantly different.  
We assume that each satellite receiver performs a separate 
time-synchronization on the desired signal.
After synchronization, since we are considering a LOS channel, it is reasonable that we
can model the channel as a single path without inter-symbol interference, at least for the 
desired signal $\bs{x}_0[k]$.  Hence, the signal $\bs{r}_m[k]$ represents the signal after time-synchronization with a time-index $k$.   

Now, after equalization, the TX signal $\bs{x}_1[k]$ may be correlated over time.  
The spatial covariance matrix $\bs{Q}_1$ in \eqref{eq:Q01} represents the
instantaneous spatial covariance.

As described in the Introduction, we assume that the $M$ satellites can 
\emph{jointly} receive signals -- for example, high-speed free-space optical
interconnects.  We can thus consider the stacked received signals and channels:
\begin{equation}
    \bs{r}[k] = \begin{bmatrix}
        \bs{r}_1[k] \\ \vdots \\ \bs{r}_M[k]
    \end{bmatrix}, \quad 
    \bs{H}_i = \begin{bmatrix}
        \bs{H}_{i,1}[k] \\ \vdots \\ \bs{H}_{i,M}[k]
    \end{bmatrix}
\end{equation}
for $i=0,1$ and write the received signal \eqref{eq:rm} as
% \begin{align}\label{IplusN}
%     \bs{r}[k] &= \bs{H}_0\bs{x}_0[k]
%     + \bs{H}_1\bs{x}_1[k]+  \bs{d}[k]
%     \nonumber \\
%     &= \bs{H}_0\bs{x}_0[k]+ \bs{v}[k],
% \end{align}
\begin{equation}
\label{IplusN} 
    \bs{r}[k] = \bs{H}_0\bs{x}_0[k]
    + \bs{H}_1\bs{x}_1[k]+  \bs{d}[k]
    = \bs{H}_0\bs{x}_0[k]+ \bs{v}[k],
\end{equation}
where $\bs{v}[k]$ is interference plus noise:
\[
    \bs{v}[k] = \bs{H}_1\bs{x}_1[k] +  \bs{d}[k].
\]
The covariance of the interference plus noise is
\begin{equation}
    \bs{P}(\bs{Q}_1) = \Exp[ \bs{v}[k]\bs{v}\herm[k]] =  \bs{H}_1\bs{Q}_1 \bs{H}_1\herm + \kappa \bs{I}. 
\end{equation}
Hence, the channel capacity after beamforming is
\begin{equation}\label{rate}
J(\bs{Q}_0, \bs{Q}_1)=\log _2 \operatorname{det}(\bs{I}+ \bs{H}_0 \bs{Q}_0 \bs{H}_0\herm \bs{P}^{-1}).
\end{equation}
We formulate the anti-jamming problem as the following min-max game where the jammer minimizes and the transmitter maximizes the achievable rate:
\begin{equation}\label{minmax}
\begin{aligned}
     \max_{\bs{Q}_0 \succeq 0} \min_{\bs{Q}_1 \succeq 0}\quad & J(\bs{Q}_0, \bs{Q}_1) \\
    \text{s.t.} \quad & \operatorname{tr}(\bs{Q}_0) \leq \mc{E}_0, \quad
         \operatorname{tr}(\bs{Q}_1) \leq \mc{E}_1.
\end{aligned}
\end{equation}

The min-max game \eqref{minmax} has an intuitive interpretation:  The 
desired and jamming transmitters are attempting to select their directions of transmissions represented through the spatial covariance matrices $\bs{Q}_0$ and $\bs{Q}_1$.
The desired transmitter ideally would concentrate energy on the strong channels (i.e., 
closest satellites).  However, it must also avoid the interference from the 
jammer.  

A critical property of \eqref{minmax} is that it admits a Nash equilibrium.
Specifically, observe that the objective function \eqref{rate} is concave in $\bs{Q}_0$
and convex in $\bs{Q}_1$.  Hence, in principle, the order of the min-max in \eqref{minmax} does not matter.  That is, the rate is the same whether the jammer ``knows" the desired transmitter's spatial covariance or the desired transmitter knows the jammer's spatial covariance matrix.

Finally, we point out that our problem formulation implicitly assumes that the desired transmitters can learn the channel of the interference $\bs{H}_1$.  This estimation may be possible since the jammer is only of concern when it is high power.  Nevertheless, the case where the jammer channel needs to be estimated is of future work.

% So, we have the following minimax problem formulation where the jammer 
% select best $\bs{Q}_1$ and the desired signal
% selects best $\bs{Q}_0$.  

% \begin{equation}
% \left\{
% \begin{aligned}
%     \wh{\bs{Q}}_1 &= \arg \min_{\bs{Q}_1} J(\wh{\bs{Q}}_0, \bs{Q}_1),\\
%     \wh{\bs{Q}}_0 &= \arg \max_{\bs{Q}_0} J(\bs{Q}_0, \wh{\bs{Q}}_1).
% \end{aligned}
% \right.
% \end{equation}

% ================= Algorithm: Best-Response + Mirror-Descent =================
\begin{algorithm}[t]
\label{algo:mirror}
\footnotesize                    
\LinesNotNumbered
\DontPrintSemicolon
\SetInd{1em}{1.15em}

\SetKwComment{tcp}{// }{}          
\SetCommentSty{textnormal}         
\SetAlgoNlRelativeSize{-1}        
\SetKwSty{footnotesize}            

\caption{Best-Response + Projected GD}
\label{alg:br-md}

\KwIn{$\bs{H}_0, \bs{H}_1, N_0, \mc{E}_0, \mc{E}_1$, step size $\eta$, tolerance $\varepsilon$}
\KwInit{$\bs{Q}_0^{(0)} = \dfrac{\mc{E}_0}{N_0}\bs{I}$, $\bs{Q}_1^{(0)} = \dfrac{\mc{E}_1}{N_1}\bs{I}$}

\For{$t = 0, 1, 2, \ldots$}{
  \tcp*[l]{Optimization of $\bs{Q}_0$}  
  $\bs{P} \gets \kappa \bs{I} + \bs{H}_1\bs{Q}_1^{(t)}\bs{H}_1\herm$

  $\bs{\Gamma} \gets \bs{H}_0\herm\bs{P}^{-1}\bs{H}_0
  = \bs{V}\bs{\Lambda}\bs{V}\herm$

  Find $\mu$ such that $\sum_i (\mu - 1/\lambda_i)_+ = \mc{E}_0$
  
  $p_i \gets (\mu - 1/\lambda_i)_+$

  $\bs{Q}_0^{(t+1)} \gets \bs{V}\,\mathrm{diag}(p_i)\,\bs{V}\herm$

  \medskip

  \tcp*[l]{Projected GD for $\bs{Q}_1$}
  $\bs{S} \gets \bs{H}_0\bs{Q}_0^{(t+1)}\bs{H}_0\herm$
  
  $\bs{G}_1 \gets \bs{H}_1\herm((\bs{P} + \bs{S})^{-1} - \bs{P}^{-1})\bs{H}_1$
  
  $\bs{Y} \gets \exp(\log \bs{Q}_1^{(t)} - \eta \bs{G}_1)$

  \vspace{2pt}
  $\bs{Q}_1^{(t+1)} \gets
  \begin{cases}
    \bs{Y}, & \text{if } \mathrm{tr}(\bs{Y}) \le \mc{E}_1,\\[3pt]
    \dfrac{\mc{E}_1}{\mathrm{tr}(\bs{Y})}\,\bs{Y}, & \text{otherwise.}
  \end{cases}$

  \medskip

  \tcp*[l]{Convergence check}
  \If{$\lVert \bs{Q}_0^{(t+1)}-\bs{Q}_0^{(t)} \rVert_{\mathrm{F}}
      + \lVert \bs{Q}_1^{(t+1)}-\bs{Q}_1^{(t)} \rVert_{\mathrm{F}} < \varepsilon$}{
    \Indp \vspace{3pt} \textbf{break} \Indm
  }
}

\BlankLine
\KwOut{$\wh{\bs{Q}}_0=\bs{Q}_0^{(t+1)}$, $\wh{\bs{Q}}_1 = \bs{Q}_1^{(t+1)}$ and rate $J(\wh{\bs{Q}}_0, \wh{\bs{Q}}_1)$}
\end{algorithm}
% ============================================================================

\section{Min-Max Optimization}\label{min-max game}

To find solutions to the min-max \eqref{minmax}, we propose a simple iterative algorithm shown in Algorithm~\ref{alg:br-md}.  The algorithm creates a sequence of estimates
$\bs{Q}^{(t)}_0, \bs{Q}^{(t)}_1$ for $t=0,1,\ldots$.
Each iteration has three steps:
\begin{enumerate}
    \item Optimization of $\bs{Q}_0$ given $\bs{Q}_1^{(t)}$ via water-filling;
    \item Projected gradient descent of $\bs{Q}_1$ given $\bs{Q}_0^{(t+1)}$;
    \item Check for convergence.
\end{enumerate}
We now describe the two optimization steps in more detail.

\medskip 
\noindent
\underline{Optimization of $\bs{Q}_0$ via water-filling}:
This step of Algorithm~\ref{alg:br-md}, seeks to find the optimal spatial covariance $\bs{Q}_0$
given the jammer's spatial covariance $\bs{Q}_1 = \bs{Q}_1^{(t)}$:
\begin{equation}
    \bs{Q}^{(t)}_0 = \arg\max_{\bs{Q}_0} J(\bs{Q}_0, \bs{Q}_1^{(t)}).
\end{equation}
%For a fixed interference covariance $\bs{Q}_1$, the receiver noise--plus--interference covariance is given in~\eqref{IplusN}.  

\begin{table}[!t]
\centering
\caption{\textbf{Simulation Parameters.}}
\label{simulation_parameters}
\renewcommand{\arraystretch}{1.1}
\large
\resizebox{0.48\textwidth}{!}{%
\begin{tabular}{|l|c|}
\hline
\textbf{Radio Parameters} & \textbf{Values} \\ \hline
Carrier Frequency and Bandwidth & \SI{10}{\giga\hertz} and \SI{100}{\mega\hertz} \\ \hline
Noise Power Spectral Density & \SI{-205}{\dBm\per\hertz} (\SI{200}{K}) \\ \hline
Desired Transmitter Power & \SI{50}{\dBm} \\ \hline
Jammer Power & \SI{70}{\dBm} \\ \hline
Atmospheric and Scintillation Attenuation & \SI{5}{\decibel} \\ \hline
\multicolumn{2}{|l|}{\textbf{Antenna Type and Configuration}} \\ \hline
Desired Transmitter & $6\times6$ URA \\ \hline
\makecell[tl]{Intelligent Jammer\\Dish Jammer} &
\makecell[t]{$6\times6$ URA\\ \SI{60}{\centi\meter} diameter} \\ \hline
Satellites & $6\times6$ URA \\ \hline
\multicolumn{2}{|l|}{\textbf{Scene Configuration}} \\ \hline
Geographical Location & (\SI{40.0822}{\degree}, \SI{-105.1092}{\degree}, \SI{1560}{\meter}) \\ \hline
Satellite Angle Range &
\makecell[t]{Azimuth: [\SI{0}{\degree}, \SI{360}{\degree}]\\ Elevation: [\SI{45}{\degree}, \SI{90}{\degree}]} \\ \hline
Satellite Constellation & Starlink (LEO) \cite{celestrak} \\ \hline
\end{tabular}%
}
\vspace{-10pt}
\end{table}

\begin{figure}[!t]
    \centering
    \includegraphics[width=0.4\textwidth]{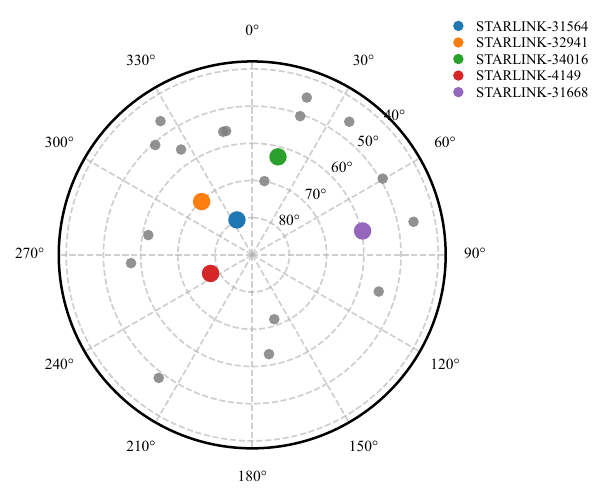}
        \vspace{-10pt}
    \caption{Instantaneous satellite geometry above the ground TX. The five nearest Starlink satellites are shown in color, and other visible ones in gray. Observation elevation: \SIrange{40}{90}{\degree}.}
    \vspace{-10pt}
    \label{nearest5sats}
\end{figure}
This maximization has a classical water-filling structure. For a fixed $\wh{Q}_1$,
the objective function \eqref{rate} is given by:
\begin{equation} \label{eq:ratea}
    J(\bs{Q}_0, \bs{Q}_1)=\log _2 \detm(\bs{I}+ \bs{Q}_0 \bs{A}).
\end{equation}
where
\begin{equation} \label{eq:Adef}
    \bs{A} = \bs{H}_0\herm\bs{P}^{-1}\bs{H}_0.
\end{equation}
Note that $\bs{A}$ is implicitly a function of $\bs{Q}_1$ via $\bs{P}$.
Now we take an eigenvalue decomposition of $\bs{A}$:
\begin{equation}
    \bs{A} = \bs{H}_0\herm\bs{P}^{-1}\bs{H}_0
      = \bs{U}\operatorname{diag}(\lambda_i)\bs{U}\herm, 
      \quad \lambda_1 \ge \!\cdots\!\ge \lambda_N \ge 0,
\end{equation}
for some unitary matrix $\bs{U}$.
Maximizing the achievable rate in~\eqref{eq:ratea}, subject to $\operatorname{tr}(\bs{Q}_0)\le \mc{E}_0$, 
the optimal $\bs{Q}_0$ follows the  water-filling structure:
\begin{equation}
    \bs{Q}_0^{(t)}
        = \bs{U}\operatorname{diag}(p_i)\bs{U}\herm,\qquad
    p_i=\!\left(\frac{1}{\mu}-\frac{1}{\lambda_i}\right)_{+},
\end{equation}
where $\mu$ satisfies $\sum_i p_i=\mc{E}_0$.

\begin{figure}[!t]
\centering
\begin{minipage}{0.93\columnwidth}
    \centering
    \includegraphics[width=\linewidth]{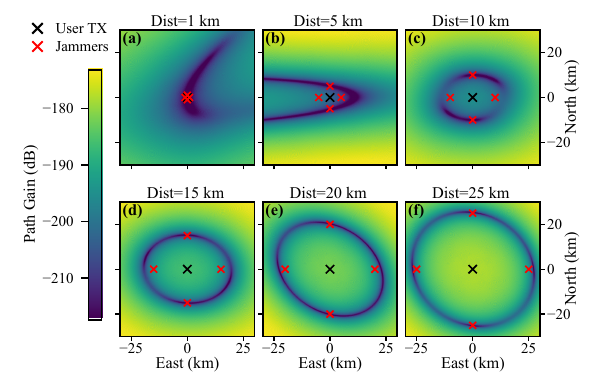}
    \vspace{-20pt}
    \caption{Capacity degradation under nullforming failure.}
    \label{nullforming_maps}
\end{minipage}
\hfill
\begin{minipage}{0.9\columnwidth}
    \centering
    \includegraphics[width=\linewidth]{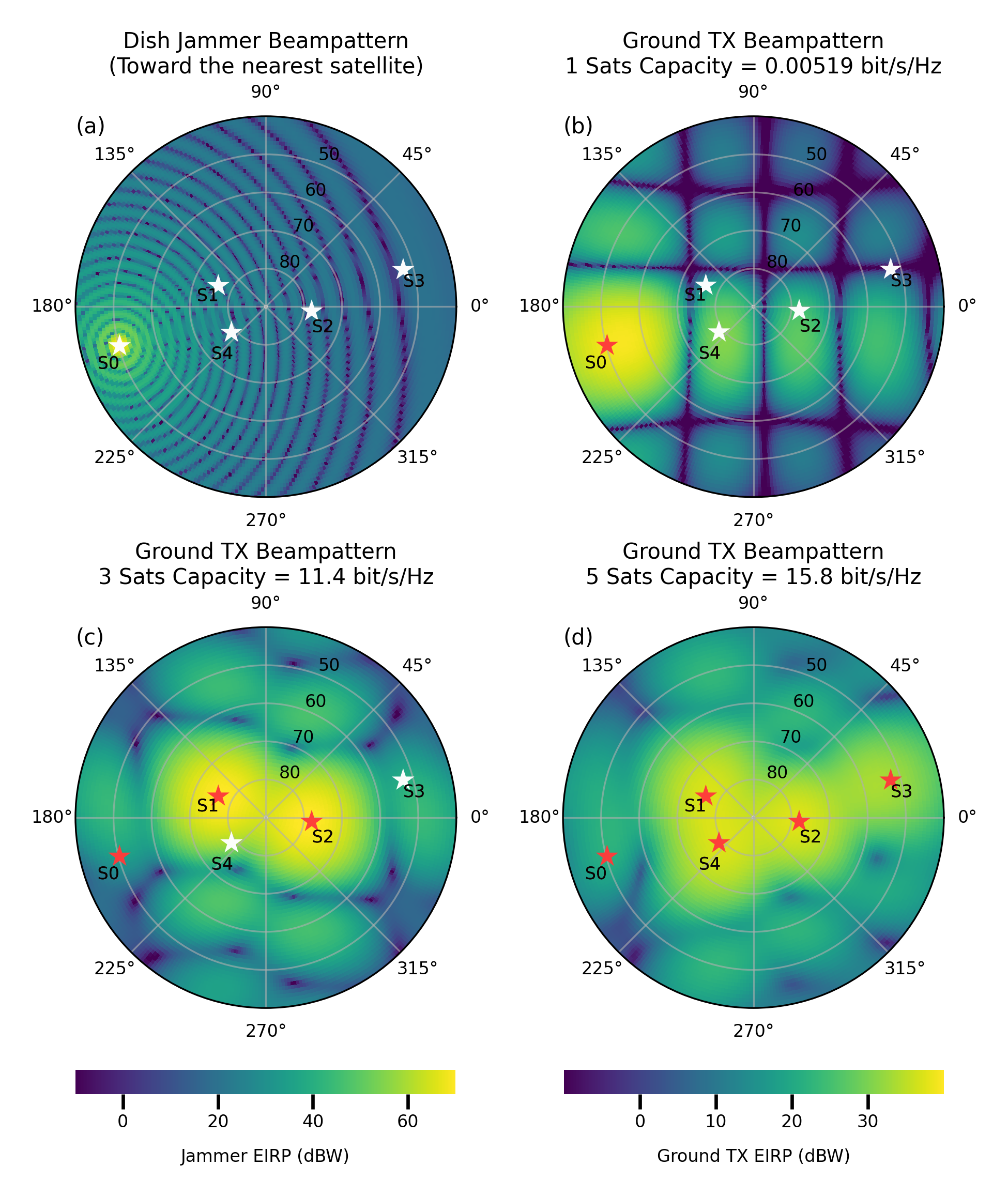}
            \vspace{-20pt}
    \caption{Beampatterns of the jammer and ground TX. Pentagram
indicate satellite positions, with red stars denoting the available
satellites. The jammer is located 1~km north of the TX, with elevation angles from 40$^\circ$ to 90$^\circ$ (zenith).}
    \label{dishpattern}
\end{minipage}\vspace{-15pt}
\end{figure}

\medskip 
\noindent
\underline{Projected gradient descent for $\bs{Q}_1$}:
The optimization of \eqref{rate} over $\bs{Q}_1$ has no closed-form solution.
We thus take a projected gradient descent step.  
For a fixed $\bs{Q}_0$, define $\bs{S}=\bs{H}_0\bs{Q}_0\bs{H}_0\herm$ and note
\begin{equation}
    J(\bs{Q}_0, \bs{Q}_1)
        = \log\!\det(\bs{P}+\bs{S})
          - \log\!\det(\bs{P}).
\end{equation}
Hence the gradient with respect to $\bs{Q}_1$ is
\begin{equation}
    \bs{G}_1 := \nabla_{\bs{Q}_1}J
        = \bs{H}_1\herm\!\big[(\bs{P}+\bs{S})^{-1}
        -\bs{P}^{-1}\big]\!\bs{H}_1 \succeq 0.
\end{equation}
The optimization in Algorithm~\ref{alg:br-md} first takes a gradient step in the direction of
the $-\bs{G}_1$.  In the update for $\bs{Y}$, an exponential is used to ensure
that the resulting matrix is positive definite.
After taking the gradient step the output matrix $\bs{Y}$ is projected to the feasible set
where $\mathrm{Tr}(\bs{Y}) \leq \mc{E}_1$.

\begin{figure}[!t]
    \centering
    \includegraphics[width=0.42\textwidth]{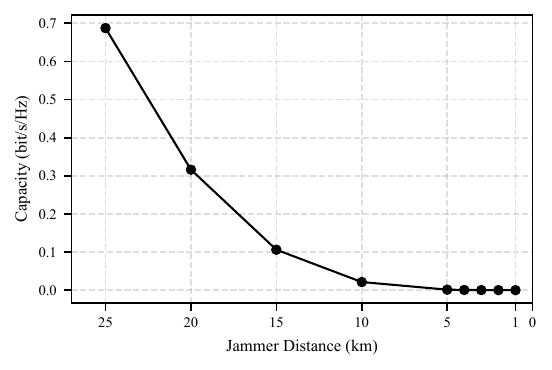}
            \vspace{-10pt}
    \caption{Capacity versus jammer distance.}
    \label{jammer_dist_c}
        \vspace{-15pt}
\end{figure}

\begin{figure}[!t]
    \centering
    \includegraphics[width=0.4\textwidth]{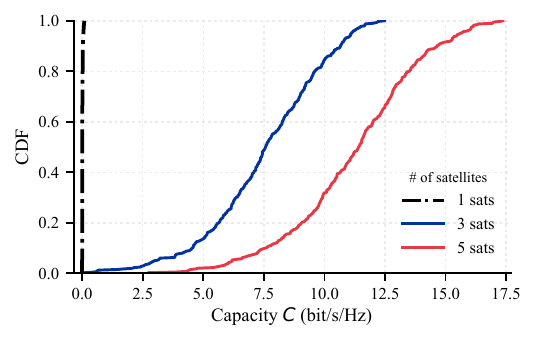}
        \vspace{-10pt}
    \caption{CDF of the achievable capacity over 400 time frames, both the dish jammer and TX utilize the nearest 1, 3, and 5 satellites. The jammer is located \SI{1}{\kilo\metre} north of the TX. }
    \vspace{-15pt}
    \label{Dishjammer CDF}
\end{figure}

\section{Practical Study and Performance Analysis}

\begin{figure*}[t]
    \centering
    \includegraphics[width=0.93\textwidth]{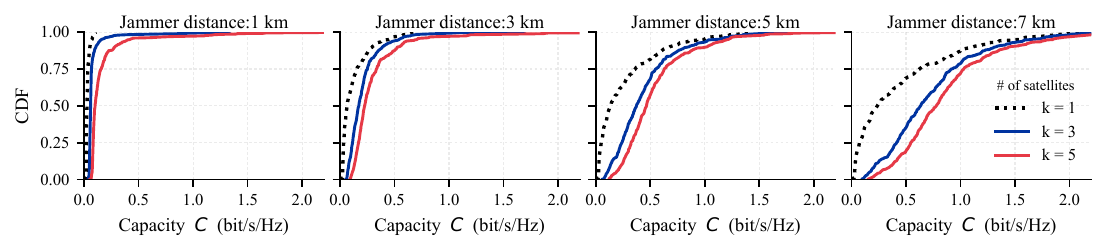}
    \vspace{-10pt}
    \caption{CDF of the achievable rate over 400 time frames, both the array jammer and TX utilize the nearest 1, 3, and 5 satellites.}
    \label{arrayjammerCDF}
        \vspace{-15pt}
\end{figure*}
\begin{figure}[!t]
    \centering
    \includegraphics[width=0.43\textwidth]{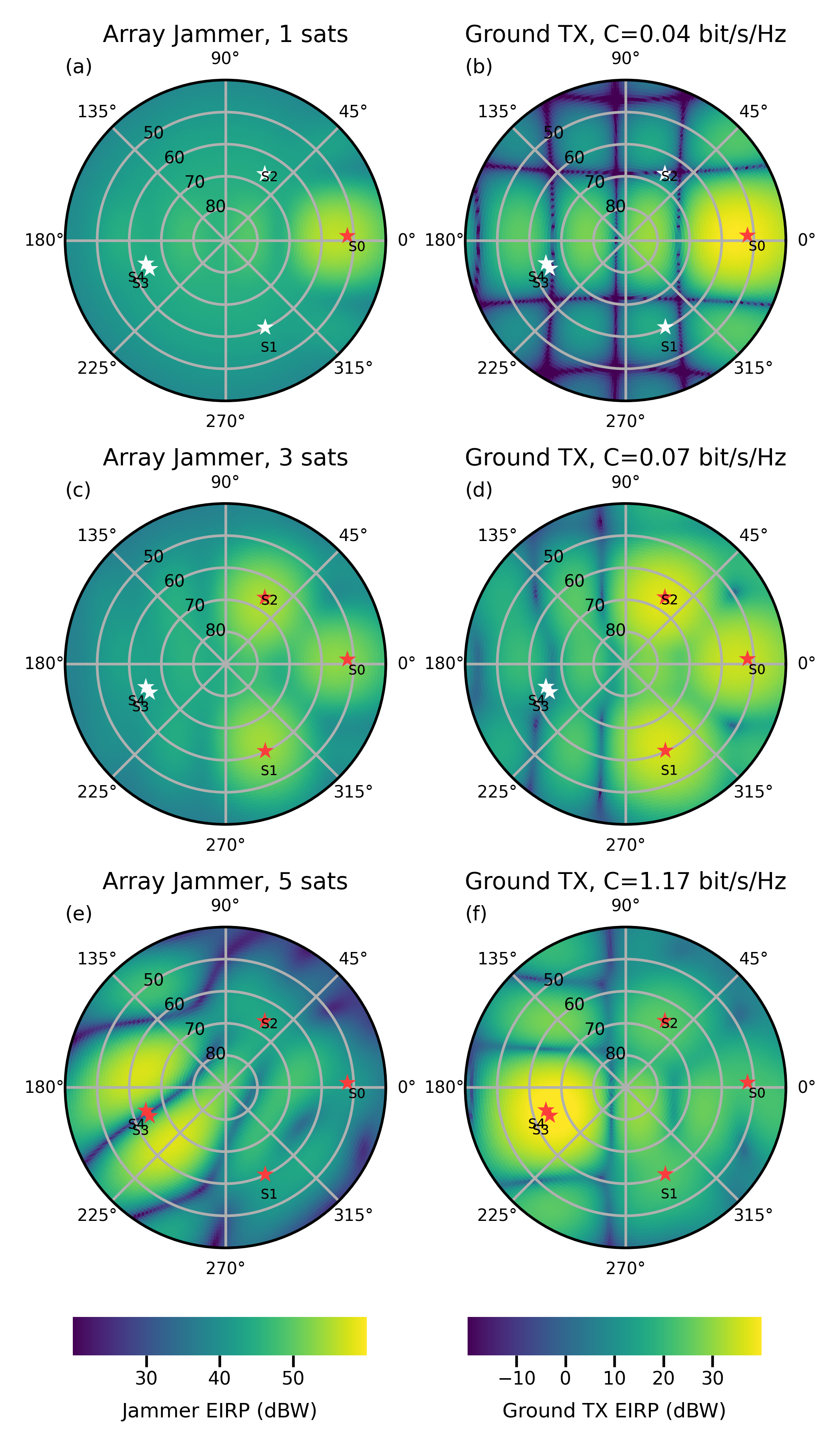}
        \vspace{-10pt}
    \caption{Beampatterns for the array jammer (left) and ground TX (right); notation as in Fig.~\ref{dishpattern}.}
    \label{arraypattern}
    \vspace{-15pt}
\end{figure}
We consider a realistic non-terrestrial communication scenario where the orbital information of the Starlink satellites is extracted from the Celestrak database~\cite{celestrak}. Among all visible satellites, the five closest ones to the ground terminal are selected as the available satellite candidates, as illustrated in Fig.~\ref{nearest5sats}. The ground transmitter (TX) is placed in a rural area near Boulder, Colorado. The antenna configurations of both the TX and the jammer are summarized in Table~\ref{simulation_parameters}. The observation angles span the full azimuth range from $0^\circ$ to $360^\circ$ and the elevation range from $40^\circ$ to $90^\circ$. The jammer's transmit power is set to be ten times higher than that of the TX, both of them are taking first 1, 3, and 5 satellites as the action space. The wireless channel is simulated using the open-source \textit{Sionna} ray-tracing platform~\cite{sionna}, taking into account realistic atmospheric absorption and scintillation attenuation effects.

\subsection{Single Satellite Nulling}
To understand the limitations of a single satellite nulling, 
consider a single nadir LEO at \SI{550}{\kilo\metre} with four high-gain dish jammers uniformly placed on a ring around the user, all boresighted to the satellite.  Fig.~\ref{nullforming_maps} shows the beam patterns on the ground when the satellite attempts to maximize the gain to desired user while placing a null at the jammers (calculations can be found in \cite{kang2024terrestrial,jia2025joint}).  It can be seen that when the ring radius is large, the satellite forms deep nulls while preserving useful gain via the interference-penalized beamformer, yielding minimal rate loss. As the jammers move inward, angular separation shrinks, nulls widen, the mainlobe distorts, and the desired gain collapses. Consequently, capacity drops rapidly with distance (Fig.~\ref{jammer_dist_c}); for separations \(\lesssim\)\SI{5}{\kilo\metre}, the link becomes effectively unusable.  These simple calculations demonstrate the limitations of anti-jamming with a single satellite.
% \begin{equation} \label{intnull}
%     \widehat{\bs{w}} \;=\; \arg \max_{\|\bs{w}\|=1} 
%     \Big[\, |\bs{w}^{*}{\bs{h}}_{0}|^{2}
%     \;-\; \lambda \sum_{i=1}^{K} |\bs{w}^{*}{\bs{h}}_{i}|^{2} \Big],
% \end{equation}
% where \({\bs{h}}_{0}\) is the desired channel, \({\bs{h}}_{i}\) are the jammer channels, and \(\lambda>0\) balances gain and interference suppression \cite{jia2025joint}.

\subsection{Dish-Type Fixed Jammer}
We next consider the case of a typical jammer with a large dish that has high power and high directional gain to jam individual satellites. This form of dish jammer, often mounted on the back of a truck, is commonly-used in many hostile scenarios.
In our framework, the jammer is TX$_1$, and we model
such a jammer as having $N_1=1$ antennas and incorporate the directionality via the antenna pattern when computing the channel matrices.
For this simulation, we consider a highly directional parabolic dish with a peak gain of \SI{40}{dBi}. 
The jammer continuously points its main lobe toward the nearest satellite to maximize interference, as illustrated in Fig.~\ref{dishpattern}(a). The ground desired transmitter (TX$_0$) employs a $6 \times 6$ uniform planar MIMO array oriented vertically upward, while the satellite antennas are oriented toward the Earth's center.

As shown in Fig.~\ref{dishpattern}(b), when the TX can access only the nearest satellite, it must beamform toward it. Due to the jammer’s much higher transmit power, the resulting capacity per second per hertz is extremely low, close to zero, as indicated by the black dashed line in Fig.~\ref{Dishjammer CDF}. However, when the TX can select the top three or five nearest satellites, it distributes its beams toward the remaining satellites. The optimized transmit covariance matrices shown in Figs.~\ref{dishpattern}(c) and~\ref{dishpattern}(d) exhibit effective ranks of two and three, respectively. As illustrated in the CDF of Fig.~\ref{Dishjammer CDF} obtained over 400 time frames, the total capacity improves significantly—more than 50\% of the cases exceed \SI{7.5}{bit/s/Hz} for three satellites, and up to \SI{11}{bit/s/Hz} for five satellites.

The results illustrate that leveraging reception on multiple satellites  with intelligent beam steering in the desired transmitters can completely mitigate the interference from 
a fixed directional jammer, even when the jammer power is significantly higher.
% \begin{figure}[!t]
%     \centering
%     \includegraphics[width=0.45\textwidth]{plot/radiomap_2x2.png}
%     \vspace{-10pt}
%     \caption{Beampatterns of the jammer and ground TX. Pentagram indicate satellite positions, with red stars denoting the available satellites. The jammer is located \SI{1}{\kilo\metre} north of the TX.}
%     \label{dishpattern}
%         \vspace{-15pt}
% \end{figure}
% \begin{figure}[!t]
%     \centering
%     \includegraphics[width=0.4\textwidth]{plot/cdf_dish_jammer.pdf}
%     \caption{CDF of the achievable capacity over 400 time frames, both the dish jammer and TX utilize the nearest 1, 3, and 5 satellites. The jammer is located \SI{1}{\kilo\metre} north of the TX. }
%     \vspace{-15pt}
%     \label{Dishjammer CDF}
% \end{figure}

\subsection{Worst-Case Jammer Scenario Analysis}
We next consider a worst-case scenario involving an intelligent array-type jammer
that can radiate multiple uncorrelated streams to different satellites. 
To this end, we assume that the jammer has a $6 \times 6$ array similar to the 
desired transmitter.
Moreover,
as a conservative assumption, we assume that the jammer can know the desired transmitter's
spatial covariance matrix and select the worst-case interference spatial covariance matrix.
In reality, estimating the desired jammer's pattern may be difficult.
Hence, our assumption is conservative.

By applying Algorithm~\ref{alg:br-md}, the jammer performs gradient-descent updates toward the equilibrium point, while in each iteration the TX executes a water-filling best response to the jammer’s strategy. With the convex–concave game formulation presented in Section~\ref{min-max game}, convergence to the equilibrium saddle point is guaranteed. We analyze one representative instance, as illustrated in Fig.~\ref{arraypattern}, where the jammer is located \SI{1}{\kilo\metre} north of the TX, representing a highly critical interference scenario. When only one or three satellites are available, as shown in Figs.~\ref{arraypattern}(a)–(d), the intelligent jammer tends to mimic the TX’s beam directions to maximize interference. Because the jammer’s transmit power is considerably higher than that of the TX, its optimal covariance matrix \(\mathrm{rank}(Q_{1}) = 36\) exhibits three dominant gain directions near the equilibrium point, whereas the TX adopts low-rank strategies with \(\mathrm{rank}(Q_{0}) = 1\) and \(\mathrm{rank}(Q_{0}) = 3\), respectively. When the number of available satellites increases to five—equivalent to an effective 180-element array on the receiver side—the TX obtains a much larger spatial strategy space. In this case, the TX employs a rank-4 transmit covariance, steering energy toward four dominant directions, while the jammer shows a slight deviation of its main beam toward satellites S3 and S4, as shown in Figs.~\ref{arraypattern}(e)–(f), which are closely aligned in angle. This expanded spatial diversity enables the TX to reach a new equilibrium point with a substantial increase in overall system capacity.

As shown in Fig.~\ref{arrayjammerCDF}, the availability of more satellites provides the TX with increased spatial degrees of freedom, leading to higher equilibrium capacities. Moreover, as the jammer moves farther from the TX, the resulting spatial channel differences further enhance system performance. Under a separation distance of \SI{7}{\kilo\metre} with five available satellites, the capacity exceeds \SI{1}{bit/s/Hz} in approximately 75\% of the cases.

\section{Conclusion}
% We addressed uplink anti-jamming for distributed LEO reception without a global timing reference by modeling the user–jammer interaction as a convex–concave min–max game in transmit covariances. Closed-form best responses lead to a simple alternating solver with mirror descent updates. Using realistic Starlink geometries, we observed sharp rate degradation at close jammer–user proximity, while access to 3–5 satellites restores spatial separability and shifts the capacity distribution notably upward for both dish and intelligent array jammers. 

This paper formulated the LEO satellite constellation uplink anti-jamming problem as a convex-concave min-max game where a ground transmitter and adversarial jammer optimize spatial transmit covariance matrices, yielding closed-form Nash equilibrium solutions through water-filling (transmitter) and eigensubspace power concentration (jammer). Sionna ray-tracing simulations with Starlink orbital geometries demonstrate that distributed reception across three to five satellites fundamentally alters the vulnerability profile compared to single-satellite links, which suffer catastrophic capacity loss when jammers approach within \SI{5}{\kilo\metre}. Three satellites achieve capacity exceeding \SI{7.5}{bit/s/Hz} in over 50\% of cases against high-power dish jammers (\SI{11}{bit/s/Hz} with five satellites), while five cooperating satellites maintain \SI{1}{bit/s/Hz} capacity in 75\% of cases at \SI{7}{\kilo\metre}
separation against intelligent array jammers with 20~dBm power advantage. Future work must address channel estimation under adversarial conditions and distributed timing synchronization across satellite networks.

\bibliographystyle{IEEEtran}
\bibliography{references}

\end{document}